\documentclass{procpavia}
\newcommand{\bra}[1]{\mbox{$\langle #1|$}}
\newcommand{\ket}[1]{\mbox{$|#1\rangle$}}

\begin{document}
\pagestyle{prochead}

\title{THE NN-FINAL-STATE INTERACTION IN TWO-NUCLEON KNOCKOUT REACTIONS}

\author{M.\ Schwamb}
\email{schwamb@kph.uni-mainz.de}
\affiliation{Dipartimento di Fisica Nucleare e Teorica
dell'Universit\`a degli Studi di  Pavia, I-27100 Pavia, Italy 
and Institut f\"ur Kernphysik,
Johannes Gutenberg-Universit\"at, D-55099 Mainz, Germany}

\author{S.\ Boffi, C.\ Giusti and F.\ D.\ Pacati}
  \email{boffi@pv.infn.it, giusti@pv.infn.it, pacati@pv.infn.it}
\affiliation
 {Dipartimento di Fisica Nucleare e Teorica dell'Universit\`a
  degli Studi di Pavia,\\ and 
	Istituto Nazionale di Fisica Nucleare, Sezione di Pavia, I-27100 
Pavia, 	Italy\\~\\}

\begin{abstract}

\noindent
 The influence  of the mutual interaction between the two outgoing
 nucleons (NN-FSI) in electro-  and photoinduced two-nucleon knockout 
 from $^{16}O$ has been  investigated perturbatively. It turns out that
 the effect of NN-FSI depends on the kinematics and on the type of reaction
 considered.   In the kinematics studied so far, the effect 
 is  larger in pp-  than in pn-knockout and in electron induced than in
 photoinduced reactions.
\end{abstract}
\maketitle
\setcounter{page}{1}


\section{Introduction}
The independent particle shell model (IPM),
 describing a nucleus as a system of 
nucleons moving in a mean field, describes many basic features of nuclear 
structure. However, the  occupation 
probabilities of various shell model orbitals
 measured via the $(e,e'p)$ reaction are in considerable disagreement with
the IPM (for an overview, see  \cite{Bof96}). Moreover, using
realistic interactions, the IPM fails to describe the binding energy of nuclei.
 This failure is a consequence of the strong short-range 
 components of the interaction, which are necessary  to reproduce NN 
data and which
 induce into the nuclear wave function correlations beyond the mean field
 description. Thus, a careful evaluation of  short-range correlations (SRC) 
 is needed to describe nuclear properties in terms of a realistic 
 NN-interaction
 and to provide profound insight into the structure of the hadronic interaction 
 in the nuclear medium \cite{MuP00}. 
SRC give an important contribution to the semi-inclusive $(e,e'p)$ reaction in
the continuum, at high values of the missing energy well beyond the two-nucleon 
emission threshold, where, however, many other competing processes contribute
and a clear identification of SRC appears very difficult. 
Thus, the most powerful and clear tool for a deeper
 investigation of SRC is the electromagnetically 
 induced  two-nucleon knockout since the probability that a real 
 or a virtual photon is absorbed by a pair of nucleons should be a direct 
 measure for the correlations between these nucleons \cite{Gott,Bof96}. 

 In general, two-nucleon knockout is a very challenging subject.
 Experimentally,  the expected  cross sections to be measured
 within a triple coincidence are exceedingly small.
 Only with the advent of high-duty-cycle 
electron beams like NIKHEF, MAMI or TJNAF a systematic investigation of
 this reaction has become possible. At present, only a few pioneering 
measurements have been carried out \cite{Ond97,Ond98,Ros00,Sta00},
 but the prospects are very encouraging \cite{MaG03,Mos03}.

 From a theoretical point
 of  view, a comprehensive treatment of the nuclear many body problem 
  of the initial state has to be performed and a profound knowledge of the
relevant reaction mechanisms is necessary. In that context, one has to be aware
of possible disturbing effects  like  contributions of two-body currents
 as well as of the final state interaction (FSI) between the two outgoing
 nucleons  and the residual nucleus, whose good understanding 
is essential to disentangle and investigate short-range effects.

 However, due to the complexity of the subject, several approximations
 have been performed in the past which restrict the reliability of the
 existing models (consider
 \cite{Bof96,Gui97,Gui98,Gui99,Gui00,Gui01,Ryc97,Ryc98a,Ryc98b}
 and the references
 therein) with respect to the interpretation of the
 experimental data.  In this context, one crucial assumption is the
 fact that the mutual interaction between the two outgoing nucleons,
 denoted as NN-FSI, can be neglected. 
Only the major contribution of FSI, due to
 the interaction of each of the two outgoing nucleons with
 the residual nucleus,
 was taken into account in the different models. The guess was
 that the effect
 of NN-FSI should not be large, at least in the superparallel kinematics,
 where the two nucleons are ejected back to back parallel and 
 antiparallel to the momentum transfer. 
 The superparallel kinematics is of particular interest for theoretical 
 \cite{Gui91} and 
 experimental \cite{Ros00,Ahr97,Arn98} investigations, 
 because in this kinematics a Rosenbluth L/T-separation makes it in principle   
 possible to extract the longitudinal structure function,
 that is assumed to be most sensitive to SRC.
 A first  calculation on nuclear matter \cite{Kno00} clearly indicates that 
 NN-FSI can in general not be neglected, even in the superparallel 
 kinematics. This result has been confirmed  by our recent work 
 \cite{ScB02,ScB03,BoG03} for two-nucleon knockout from a complex nucleus 
 like $^{16}O$ within the unfactorized approach of \cite{Gui97}.
  The corresponding theoretical framework
 and the adopted approximations are outlined in section \ref{mod}.  Numerical
 results for some selected kinematical situations are presented in
 section \ref{res}.  Some perspectives of possible improvements 
  and future developments are given in section \ref{sum}.

\section{The model}\label{mod}
The cross section for electromagnetic two-nucleon-knockout is given in
general by the square of the scalar product of the relativistic
electron current $j^{\mu}$ and of the nuclear current $J^{\mu}$, where
the latter is given by the Fourier transform of the transition matrix
element of the charge-current density operator between initial and final
 nuclear states, i.e., 
\begin{equation}\label{eq1}
J^{\mu}(\vec{q}\,) = \int \bra{\Psi_f} \hat{J}^{\mu}(r) \ket{\Psi_i}
e^{i \vec{q} \cdot \vec{r}} {\mathrm d}\vec{r} \,\, .
\end{equation}
The model is based on  the two assumptions of an exclusive reaction, for the 
transition to a specific discrete state of the residual nucleus, and 
of the direct knockout mechanism, i.e., we assume a direct one-step process
where the photon directly interacts with the pair of nucleons that are emitted 
and the A-2 
nucleons of the residual nucleus behave as spectators.

As a result of these two assumptions, the integral (\ref{eq1}) 
can be reduced to a form with three main ingredients: the 
two-nucleon overlap function (TOF) between the ground state of the target 
and the final state of the residual nucleus, the nuclear
 current $\hat{j}^{\mu}$ of the two emitted nucleons,
 and the two-nucleon scattering wave 
 function $\ket{\psi_f}$.

\begin{figure}[b]
\centerline{\includegraphics[width=8cm,angle=0]{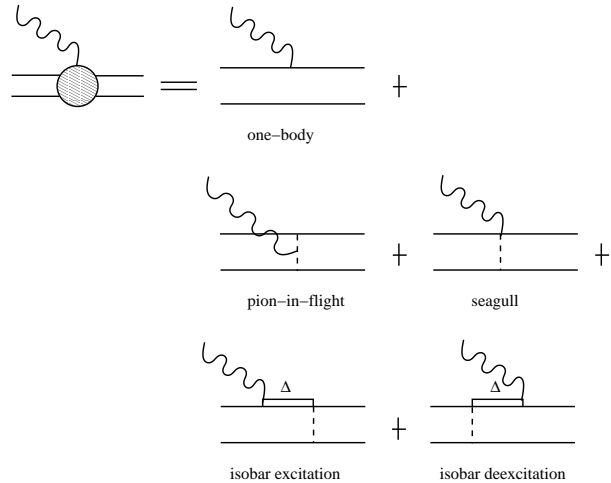}}
\vspace{0.2cm}
\caption{The electromagnetic current contributions taken into account
in the present approach.}
\label{fig_strom}
\end{figure}

In general, the nuclear current operator $\hat{j}^{\mu}(r)$ is the sum of 
 a one-body and a two-body part (Fig.\  \ref{fig_strom}). 
 The one-body part consists of  the usual  charge operator and the convection 
and spin currents. The two-body part 
consists of the  nonrelativistic  pionic seagull MEC, the pion-in-flight MEC 
and the $\Delta$-contribution. For more details see  \cite{Gui98,WiA97}. 

The TOF requires a calculation of the two-hole spectral
function including consistently different types of correlations, i.e. SRC
and tensor correlations (TC), as well as long-range correlations (LRC),
 which are mainly representing
collective excitations of nucleons at the nuclear surface.  SRC
 and TC 
are introduced in the radial wave function of the relative
motion by means  of state dependent defect functions which are added to the 
uncorrelated partial wave. For the pp-case \cite{Gui97,Geu96}, the defect 
functions are obtained by solving the Bethe-Goldstone equation using
 the  Bonn OBEPQ-A potential \cite{Mac89}.
 For the pn-case \cite{Gui99}, 
SRC and TC correlations are calculated within the framework of the 
coupled-cluster method  \cite{MuP00} with the AV14-potential \cite{WiS84} and 
using the so-called $S_2$ approximation, where only $1$-particle $1$-hole and 
$2$-particle $2$-hole excitations are included in the correlation
operator.  This method is an extension of the Bethe-Goldstone equation and  
takes into account, among other things and  besides particle-particle ladders, 
also hole-hole ladders. These, however, turn out to be rather 
 small  in $^{16}O$ \cite{MuP00}, so that the two approaches are similar in 
 the treatment of SRC.
LRC are included in the expansion coefficients of the TOF. For the pp-case, 
these coefficients are calculated in an extended shell-model basis within a dressed 
random phase approximation \cite{Gui97,Geu96}. For the pn-case, a simple 
configuration mixing calculation of the two-hole
 states in $^{16}O$ has been done and only $1p$-hole states are considered for
 transitions  to the low-lying states of $^{14}N$ \cite{Gui99}.

\begin{figure}[t]
\centerline{\includegraphics[width=16cm,angle=0]{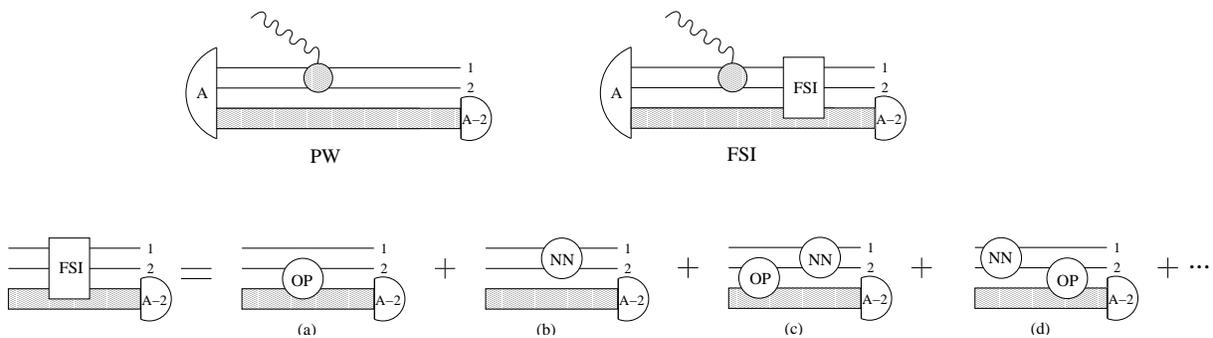}}
\vspace{0.2cm}
\caption{The relevant diagrams for electromagnetic 
 two-nucleon knockout on a complex nucleus
A. The two diagrams on top depict the plane-wave approximation 
 (PW)  and  the distortion of the two outgoing nucleon wave functions
by final state interactions (FSI).  Below, the relevant mechanisms of 
FSI are depicted in detail, where the open circle denotes either the  
 nucleon-nucleus scattering amplitude (OP), or the
nucleon nucleon-scattering amplitude  (NN), see equation
 (\protect{\ref{tnn}}).
Diagrams which are given by an interchange of nucleon 1 and 2 are not depicted.
In the present approach, only diagrams (a) and (b) are taken into account.
}
\label{fig_mechanism}
\end{figure}

Concerning the treatment of the final state two-nucleon scattering function
 $\ket{\psi_f}$, we will discuss different approximations frequently 
 used in the past. 
In the simplest approach any interaction between the two nucleons and the
 residual nucleus is neglected  and a plane-wave (PW) approximation is assumed
  for the two outgoing nucleons.  
 In the more sophisticated
 approach of \cite{Gui97} (DW),  the interaction between each of the outgoing 
 nucleons and the residual nucleus is considered by using a complex 
 phenomenological optical potential $V^{OP}$ for nucleon-nucleus scattering  
 which contains a central, a Coulomb and a spin-orbit term \cite{Nad81} (see 
 diagram (a) in Fig.\ \ref{fig_mechanism}). Only very recently the mutual 
NN-interaction  between the two outgoing nucleons, called NN-FSI, has been 
taken into account \cite{ScB02,ScB03,BoG03}, see diagram (b) in Fig.\
 \ref{fig_mechanism}.  Multiscattering processes like those depicted by 
diagrams (c) and (d) of  Fig.\ \ref{fig_mechanism} would require a 
 genuine three-body approach which is presently under investigation. At the 
 moment, these diagrams are neglected so that the results discussed in the
 next section must be considered as preliminary. The hope of the present 
 approximation is to obtain a first reliable estimate of the role of NN-FSI 
 in various kinematical situations for two-nucleon knockout from $^{16}O$.
 The present treatment of incorporating FSI  due both to 
the nucleon-nucleus interaction  $V^{OP}$
 and to the mutual NN-interaction $V^{NN}$ is denoted by DW-NN.
In addition, we denote  PW-NN the treatment where only $V^{NN}$ is 
 considered and $V^{OP}$ is switched off. 

In a more quantitative form, denoting  $\ket{\vec{p}^{\,\,0}_i}$
 a plane-wave state of nucleon $i$, with momentum $\vec{p}^{\,\,0}_i$,
 and by  $\ket{\phi^{OP}(\vec{p}^{\,\,0}_i)}$ the state of nucleon $i$ 
 distorted by the optical potential according to the 
Schr\"o\-dinger equation
\begin{equation}\label{op-equation}
\left( H_0(i) + V^{OP}(i) \right) \ket{\phi^{OP}(\vec{p}^{\,\,0}_i)} = E_i 
\ket{\phi^{OP}(\vec{p}^{\,\,0}_i)},
\end{equation}
 where $H_0(i)$ denotes the kinetic energy operator, the corresponding
 final states in these different approximations are given by
\begin{eqnarray}
 \ket{\psi_f}^{PW} &=&
 \ket{\vec{p}^{\,\,0}_1}\,\ket{\vec{p}^{\,\,0}_2}\,\, , \label{fsi-approx1} \\
 \ket{\psi_f}^{DW} &=&
 \ket{\phi^{OP}(\vec{p}^{\,\,0}_1)}\,\ket{\phi^{OP}(\vec{p}^{\,\,0}_2)}
\,\, ,  \label{fsi-approx2} \\
\ket{\psi_f}^{PW-NN} &=& \ket{\vec{p}^{\,\,0}_1}\,\ket{\vec{p}^{\,\,0}_2}
+  G_0(z) T^{NN}(z) 
 \ket{\vec{p}^{\,\,0}_1}\,\ket{\vec{p}^{\,\,0}_2}\,\,. \label{fsi-approx3} 
\end{eqnarray}
In the last equation, the NN-scattering amplitude $T^{NN}$ is given by
 ($z= \frac{(\vec{p}^{\,0}_1)^2}{2m_p} + \frac{(\vec{p}^{\,0}_2)^2}{2m_p} 
+ i \epsilon$)
\begin{equation}\label{tnn}
T^{NN}(z) = V^{NN} + V^{NN} G_0(z) T^{NN}(z), 
\end{equation}
with the free propagator
\begin{equation}\label{g0}
G_0(z) = \frac{1}{z-H_0(1)-H_0(2)}.
\end{equation}
Moreover, our full approach DW-NN is given by 
\begin{equation}\label{fsi-approx4}
 \ket{\psi_f}^{DW-NN} = \ket{\phi^{OP}(\vec{p}^{\,\,0}_1)}\,
\ket{\phi^{OP}(\vec{p}^{\,\,0}_2)} + G_0(z) T^{NN}(z) 
\ket{\vec{p}^{\,\,0}_1}\,\ket{\vec{p}^{\,\,0}_2}.
\end{equation}
 In our practical calculations, the NN-scattering amplitude $T^{NN}$
 is evaluated with the help of a usual partial wave decomposition of 
 the NN-interaction $V^{NN}$ taking into account all contributions
 up to an orbital angular momentum of 3, i.e., the isospin-1 partial waves
 $^1\!S_0, ^3\!P_0, ^3\!P_1, ^3\!P_2, ^1\!D_2, ^3\!F_2, ^3\!F_3$,
 and $^3\!F_4$  contributions 
 for pp-knockout and moreover the isospin-0 contributions
 $^3\!S_1, ^1\!P_1,  ^3\!D_1, ^3\!D_2,  ^3\!D_3$, and $^1\!F_3$ 
 for pn-knockout. It has been checked  that this truncation
 is sufficient at least   for the kinematics considered
 in this paper. More details concerning the numerical implementation
 of the NN-FSI can be found in \cite{ScB03}. 

\section{Results}\label{res}

\begin{figure}[p]
\centerline{\includegraphics[width=14cm,angle=0]{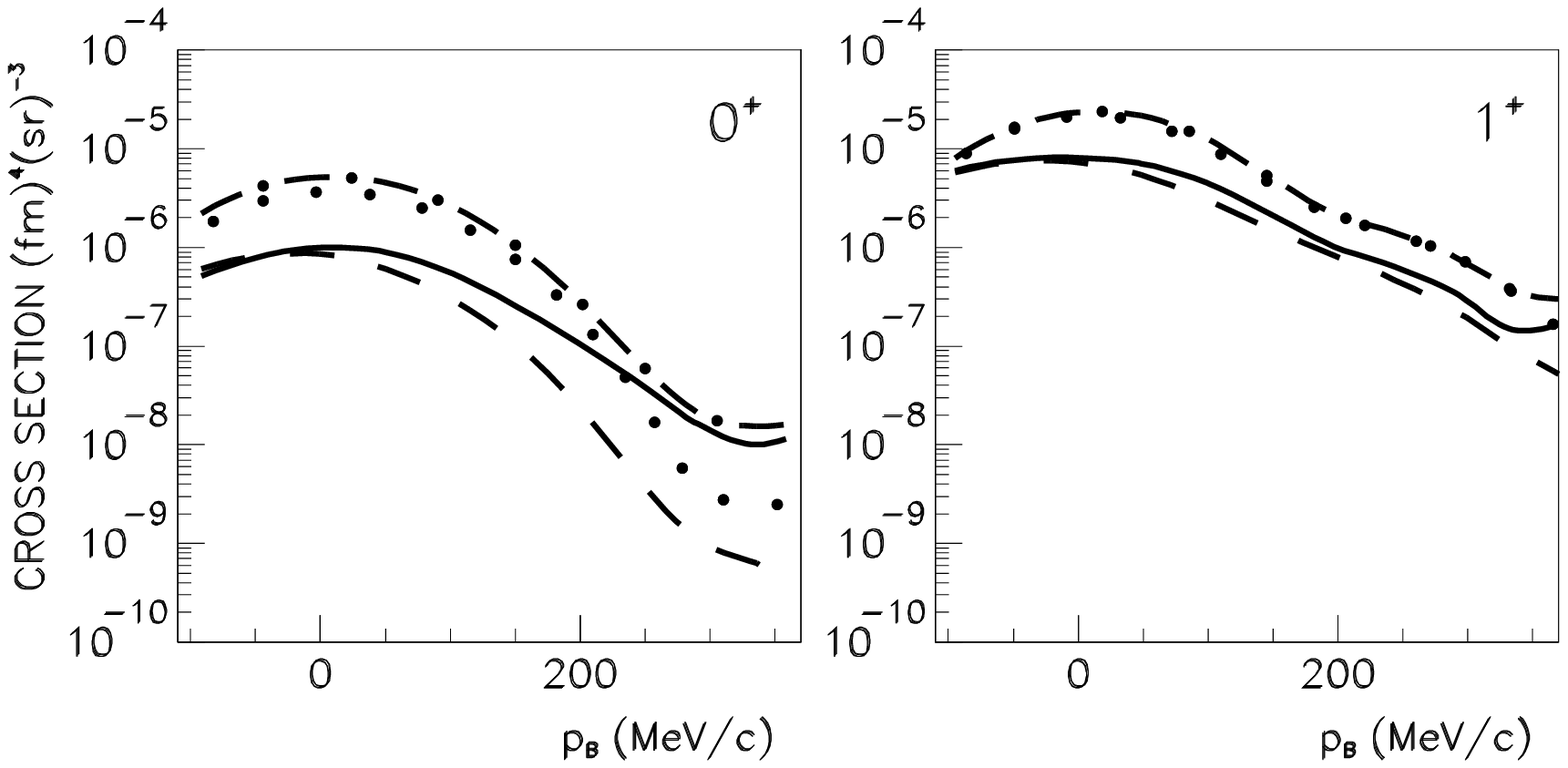}}
\vspace{0.2cm}
\caption{The differential cross section of the $^{16}O(e,e' pp)$ reaction to 
the $0^+$ ground state of $^{14}C$ (left panel) and of the $^{16}O(e,e' pn)$ 
reaction to the $1^+$ ground state of $^{14}N$ (right panel) in a  
superparallel kinematics  with an incident electron energy $E_0= 855$ MeV,  
an electron scattering angle $\theta_e = 18^{\circ}$, energy transfer 
$\omega=215$ MeV and  $q=316$ MeV/$c$.
In $^{16}O(e,e' pn)$ the proton is ejected parallel and the neutron 
antiparallel to $\vec{q}$.  
 Different values of $p_{\rm B}$ are obtained changing the kinetic energies of the outgoing
nucleons. Positive (negative) values of $p_{\rm B}$ refer to situations where 
${\vec p}_{\rm B}$ is parallel (anti-parallel) to ${\vec q}$.
 Line convention:
 PW (dotted), PW-NN (dash-dotted), DW (dashed), DW-NN (solid).
\label{result1}
}

\vspace{0.2cm}

\centerline{\includegraphics[width=9cm,angle=0]{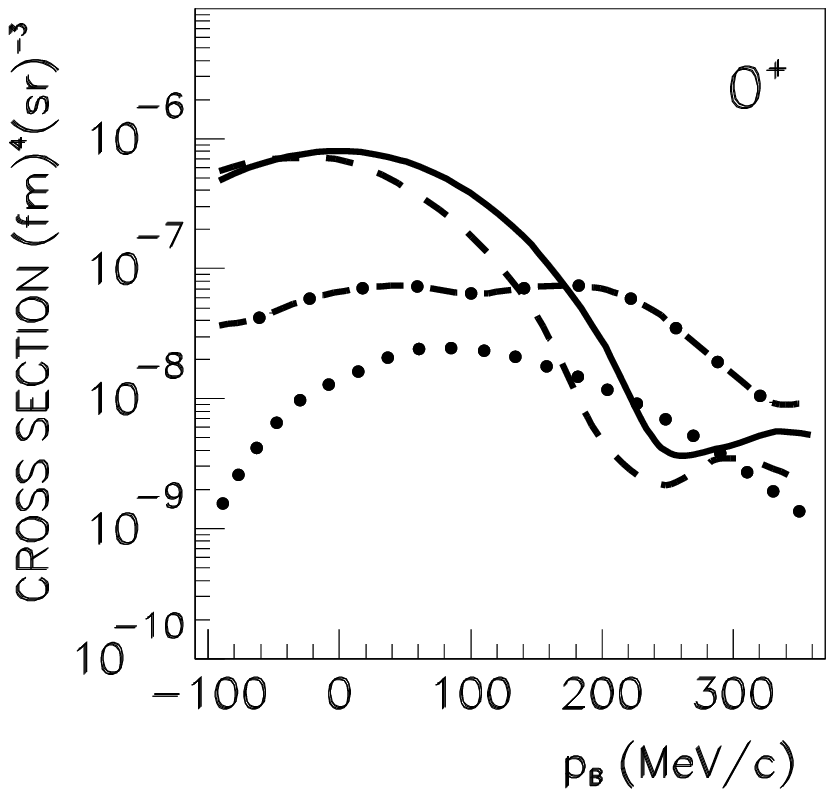}}
\vspace{0.5cm}
\caption{The differential cross section of the $^{16}O(e,e' pp)$ reaction to 
the $0^+$ ground state of $^{14}C$
 in the same superparallel kinematics as in Fig.\ \protect{\ref{result1}}.
Line convention:
 DW with the $\Delta$-current (dotted), 
 DW-NN with the $\Delta$-current (dash-dotted),
 DW with the one-body current (dashed),
 DW-NN with the one-body current (solid).
\label{result2}
}

\end{figure}

In this section, we discuss the role of NN-FSI on different electromagnetic
reactions in various kinematics.
We start with the superparallel kinematics of a recent Mainz experiment 
 \cite{Ros00}.
The differential cross sections calculated within the different
 approximations  (\ref{fsi-approx1}-\ref{fsi-approx3}, \ref{fsi-approx4}) for 
 the $^{16}O(e,e'pp)$ reaction to
 the  $0^+$ ground state of $^{14}C$ and the $^{16}O(e,e'pn)$ reaction to 
 the $1^+$ ground state of $^{14}N$   
are displayed in the left and right panels
 of Fig.\ \ref{result1}, respectively.
It can be clearly seen in the figure that the inclusion of the 
optical potential leads, in both reactions, to an overall and substantial
 reduction of the
calculated cross sections (see the difference between the PW and DW results),
 which, e.g. at $p_{\rm B}=100$ MeV/$c$, corresponds to a factor of $\sim$
 0.2 in $(e,e'pp)$ and of $\sim$ 0.3 in $(e,e'pn)$.
This effect is well known and it is mainly due to the imaginary part of 
the optical potential, that accounts for the flux lost to inelastic channels in
the nucleon-residual nucleus elastic scattering. 
The optical potential gives the
dominant contribution of FSI for recoil-momentum values up to  
$p_{\rm B} \simeq 150$ MeV/$c$. At larger values NN-FSI gives an 
enhancement of 
the cross section, that increases with $p_{\rm B}$. In $(e,e'pp)$ this 
enhancement 
goes beyond the PW result and amounts to roughly an order of magnitude for 
$p_{\rm B} \simeq 300$ MeV/$c$. In $(e,e'pn)$ this effect is still sizeable
 but  much weaker, i.e. only 50$\%$ of enhancement at $p_{\rm B}=300$ 
 MeV/$c$.

It can be seen from Fig.\  \ref{result2}, 
that the substantial increase of the cross section
in pp-knockout at large missing momenta is mainly
 due to a strong enhancement of the $\Delta$-current contribution by NN-FSI.
Up to about 100-150 MeV/$c$, 
however, this effect is completely overwhelmed by the dominant contribution 
of the one-body current, while for larger values of $p_{\rm B}$,
 where the one-body 
current is less important in the cross section, the increase of the 
$\Delta$-current is responsible for the substantial enhancement in the final 
result of Fig.\ \ref{result1}. 
The effect of NN-FSI on the one-body current is  anyhow 
 sizeable (a  factor of about 2 at $p_{\rm B}=100$ MeV/$c$),
 and it is responsible for the NN-FSI effect at lower  and intermediate
 values of   $p_{\rm B}$ in Fig.\ \ref{result1}.

As has been outlined in \cite{ScB03}, the main difference concerning the 
role of NN-FSI in pp- and pn-emission is that in the latter
reaction  the seagull current, which is not present
 in pp-emission,  is also largely affected by NN-FSI. It turns out that
 the strong effects of NN-FSI on the $\Delta$-contribution
  and on the seagull current
tend to cancel each other to a large extent, so that the 
corresponding pn-cross section is only moderately affected.
It should be stressed that we have found such a cancellation in pn-emission
 only for the unpolarized cross 
section in the superparallel kinematics, which is, however, a situation of 
particular experimental interest. We did not check till now if a similar
cancellation occurs for other kinematics or in polarization observables.

\begin{figure}[t]
\centerline{\includegraphics[width=8cm,angle=0]{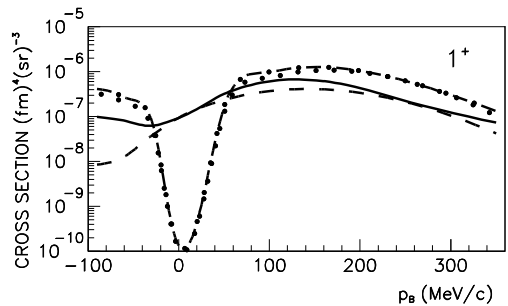}}
\caption{The differential cross section  in the $^{16}\mbox{O}(e,e' pp)$
  reaction
to the $1^+$  state of $^{14}\mbox{C}$
  for the superparallel kinematics discussed
 in the text. Line notation as in Fig.\  \ref{result1}.
}
\label{fig1+}
\end{figure}

 Another important result of our studies is that the role of NN-FSI
 depends strongly of the specific state of the residual nucleus. 
 In general, for the kinematics studied so far in pp-knockout,
 it turns out that the $^1S_0$ relative state of the pp-pair in the target is 
 more affected by NN-FSI than the  higher partial waves. 
 It is known from previous work \cite{Gui97} that the $^1S_0$ state dominates
 the $^{16}\mbox{O}(e,e' pp)$ cross section to the $0^+$ ground state of 
 $^{14}\mbox{C}$. Thus, the strong effect of NN-FSI found for this transition 
 is not surprising. On the 
other hand,   for the transition to the $1^+$ excited state, where  
 the  $^1S_0$ partial wave  in the initial state cannot  contribute,
 it is shown in Fig.\  \ref{fig1+} that 
the effect of NN-FSI becomes considerably smaller, but anyhow not negligible.

\begin{figure}[!pht]
\centerline{\includegraphics[width=8cm,angle=0]{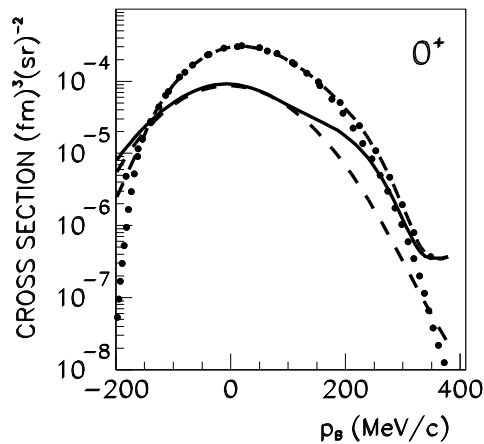}}
\vspace{0.2cm}
\caption{The differential cross section of the $^{16}O(\gamma,pp)$ reaction to 
the $0^+$ ground state of $^{14}C$ in superparallel kinematics at 
$E_\gamma =$ 215 MeV. 
 Line convention  as in Fig.\  \protect{\ref{result1}}.
\label{result4}
}
\end{figure}

\begin{figure}[!ht]
\centerline{\includegraphics[width=8cm,angle=0]{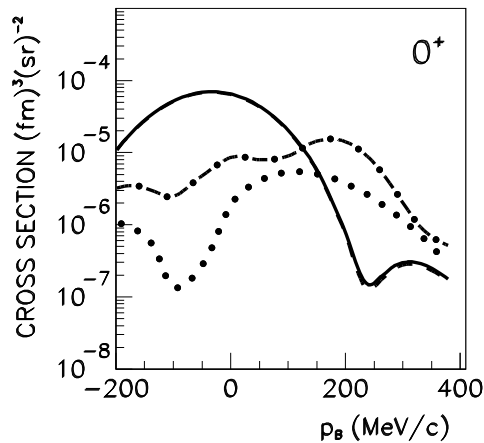}}
\vspace{0.2cm}
\caption{The differential cross section of the $^{16}O(\gamma,pp)$ reaction to 
the $0^+$ ground state of $^{14}C$ in the same kinematics as in 
Fig.\ \protect{\ref{result4}}. 
 Line convention  as in 
 Fig.\  \protect{\ref{result2}}. 
\label{result5}
}
\end{figure}

At next, we turn to photoinduced pp-knockout.
The cross section of the $^{16}O(\gamma,pp)$ reaction to the  $0^+$ ground 
state of $^{14}C$  calculated with the different approximations for FSI 
is shown in Fig.\ \ref{result4}. 
The separated contributions of the one-body 
and $\Delta$-currents in DW and DW-NN  are displayed in Fig.\ \ref{result5}. 
Calculations have been performed in superparallel kinematics, and for an
incident photon  energy which has the same value, $E_\gamma= 215$  MeV, as 
the energy transfer in the $(e,e'pp)$ calculation of Fig.\ \ref{result1}. 
This kinematics, which is not very well 
suited for $(\gamma,pp)$ experiments, is interesting for a
theoretical comparison with the corresponding results of the electron induced
reaction in Figs.\ \ref{result1} and \ref{result2}.
In general, two-body currents give the major contribution to $(\gamma,NN)$ 
reactions. In this superparallel kinematics, however, the  $(\gamma,pp)$ 
cross section is dominated by the one-body current for recoil momentum values 
up to about 150 MeV/$c$. For larger values the $\Delta$-current plays the main
role. This is the same behavior as in the corresponding situation for 
$(e,e'pp)$.   Similar to $(e,e'pp)$,  NN-FSI produces an 
enhancement of the $\Delta$-current contribution, see Fig.\ \ref{result5},
 whose absolute size strongly depends on  $p_{\rm B}$. Whereas for
 $p_{\rm B}=50$ MeV/$c$, the effect is only $\sim 70 \%$, one obtains
 more than one order of magnitude of enhancement at $p_{\rm B}=-100$ MeV/$c$.
 The role 
of NN-FSI on the one-body current is practically negligible
 in $(\gamma,pp)$, while it is significant in  $(e,e'pp)$ as has been 
 discussed above. 
 This effect is produced in $(e,e'pp)$ on
the longitudinal part of the nuclear current, that does not contribute
 in reactions induced by a real photon. 
Thus, in practice, in this kinematics NN-FSI affects only the $\Delta$-current 
and therefore in Fig.\ \ref{result4} its effect is negligible in the region 
where the one-body current is dominant. At large values of $p_{\rm B}$, where 
the role of the $\Delta$-current becomes important, the enhancement produced 
by NN-FSI is   large, i.e.\ a factor of $\sim$ 4 at  $p_{\rm B}=300$ 
 MeV/$c$, 
 but  nevertheless   weaker
than in the same superparallel kinematics for $(e,e'pp)$.


\begin{figure}[!pht]
\vspace{0.2cm}
\centerline{\includegraphics[width=14cm,angle=0]{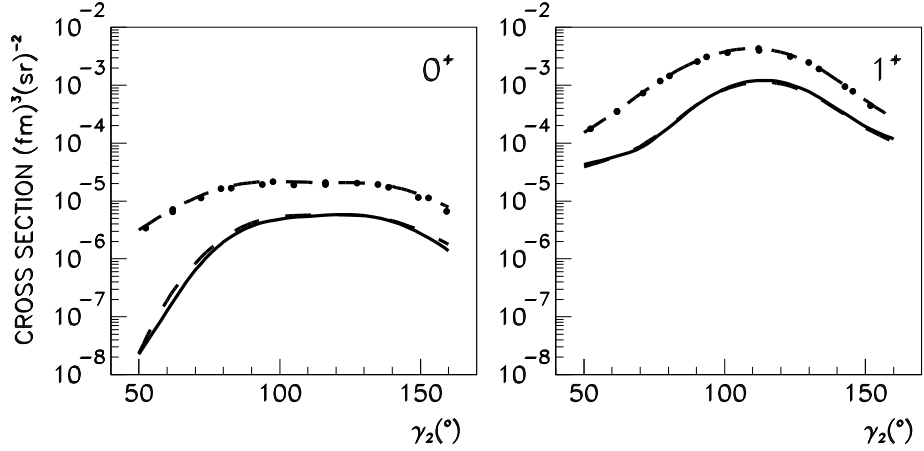}}
\caption{The differential cross section  of 
  the $^{16}O(\gamma, pp)$ reaction to the $0^+$ ground state of $^{14}C$ 
  (left panel) and of the $^{16}O(\gamma, pn)$ reaction to the $1^+$ ground 
  state of $^{14}N$ (right panel)  as a function of the scattering angle 
  $\gamma_2$ of the second outgoing nucleon in a coplanar kinematics with 
  $E_\gamma =$ 120 MeV, $T_1 = 45$ MeV and $\gamma_1=45^{\circ}$. 
  Line convention as in Fig.\  \protect{\ref{result1}}.
\label{result6}
 }
\end{figure}

\begin{figure}[!ht]
\vspace{0.2cm}
\centerline{\includegraphics[width=7cm,angle=0]{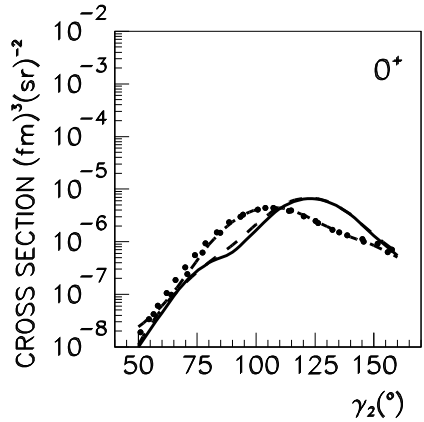}}
\caption{The differential cross section  of 
  the $^{16}O(\gamma, pp)$ reaction to the $0^+$ ground state of $^{14}C$ 
  in the same kinematics as in Fig.\ \protect{\ref{result6}}.
  Line convention as in Fig.\ \protect{\ref{result2}}.
\label{result7}
  }
\end{figure}

Another example is presented in Fig.\ \ref{result6}, where the results of the
different approximations in the treatment of FSI are displayed for the
  $^{16}O(\gamma,pp)$ reaction to the  $0^+$ ground state of $^{14}C$ 
(left panel)
and for the $^{16}O(\gamma,pn)$ reaction to  the $1^+$ ground state
 of $^{14}N$ 
(right panel) in a coplanar kinematics at $E_\gamma = 120$ MeV, where the 
energy and the scattering angle of the outgoing proton are fixed at 
$T_1= 45$ MeV and $\gamma_1= 45^{\circ}$, respectively. Different values of 
the recoil momentum  can be obtained by varying the scattering angle 
$\gamma_2$ 
of the second outgoing nucleon on the other side of the
 photon momentum.  It can be clearly seen in the figure 
that NN-FSI has almost no effect. In contrast, a very large contribution is 
given, for both reactions, by the optical potential, which produces again a 
substantial reduction.
This kinematics, which appears within reach of available experimental
facilities,  was already envisaged in \cite{Gui01} as promising to study SRC in
the $(\gamma,pp)$ reaction. In fact, at the considered value of the photon
energy, the contribution of the $\Delta$-current is relatively much less 
important, and while the $(\gamma,pn)$ cross section is dominated by the 
seagull current \cite{Gui01}, in the $(\gamma,pp)$ cross section the 
contribution of the one-body current is large and competitive with the one of 
the two-body current. This can be seen in Fig.\ \ref{result7}, 
where the two separated contributions are shown in the DW and  in the DW-NN
approximations. Both processes are important: the $\Delta$-current plays the
main role at lower values of $\gamma_2$, while for  $\gamma_2 \geq 110^{\circ}$
the one-body current and therefore SRC give the major contribution. The effect
of NN-FSI is practically negligible on both terms, which explains the result in
the final cross section of Fig.\ \ref{result6}.
A study of the $(\gamma,pp)$ reaction in a kinematics of the type considered
 in Figs.\ \ref{result6} and \ref{result7}, where NN-FSI is negligible and
 correlations are important, might represent a promising alternative to the 
 $(e,e'pp)$ reaction for the investigation of SRC.


\section{Summary and Outlook}
\label{sum}
Exclusive experiments with direct two-nucleon emission by an electromagnetic
 probe have been suggested a long time ago as good candidates to study 
correlations beyond a mean field description of nuclei.
 The study of these reactions is very challenging both for experiment and
 theory. Only the presently available high-duty cycle accelerators
 allow the corresponding measurements of the exceedingly small cross sections.
 First experiments have already been performed or are presently under 
investigation.  From a theoretical point of view, a good understanding of the 
 relevant reaction mechanisms is necessary for the interpretation
 of the data. In that context, disturbing effects like the role of two-body
 currents or final state interactions (FSI)  must be well under control. 
 Concerning FSI, a consistent evaluation would require a three-body
 approach for the two nucleons and the residual nucleus. So far, only
 the major contribution of FSI, 
 due to the interaction of each of 
the two outgoing nucleons with the residual nucleus,
 was taken into account in the different models. The original guess 
was that the 
 mutual interaction between the two outgoing nucleons (NN-FSI) could be
 neglected.

 In the present work, we have studied the role of  NN-FSI 
within a perturbative treatment which should give a first reliable idea of 
their  relevance. It turns out that  NN-FSI are in general
 not negligible. 
 However, their absolute size strongly depends on the chosen kinematics,
 on the type of reaction and on the final state of the residual nucleus. 
  In the kinematics studied till now, NN-FSI effects are in general larger
 in pp- than in pn-knockout and in electro- than in photoinduced reactions. 
 They affect in a different manner the various terms of the nuclear current,
 usually more the two-body than the one-body terms, and they are 
 sensitive to the various theoretical ingredients of the reaction. This makes
 it difficult to make predictions about the role of NN-FSI in a particular
 situation. Each specific situation should be individually investigated.
 
In order to improve the reliability of the theoretical description of the
two-nucleon knockout process, the full three-body problem of the final state has
to be tackled in forthcoming studies. In that context, special emphasis has to
be devoted to a more consistent treatment of the initial and the final state.

\vspace{0.2cm} 
\centerline{{\bf Acknowledgements}}
\vspace{0.2cm}
This work has partly been performed under the contract  HPRN-CT-2000-00130 of
the European Commission. Moreover, it has been supported  by 
 the Istituto Na\-zionale di Fisica Nucleare and by the
 Deutsche For\-schungs\-gemein\-schaft (SFB 443). Fruitful discussions with
 H.\ Arenh\"ovel are gratefully acknowledged.

\end{document}